%% file: main.tex
\documentclass[aps,prl,floatfix,reprint,showpacs,superscriptaddress,groupedaddress]{revtex4-1}
\usepackage{latexsym}
\usepackage{graphicx}
\usepackage{rotating}
\usepackage[normalem]{ulem}
\usepackage{hyperref}
\usepackage{amsmath,amssymb,amsfonts}
\usepackage{bm}
\usepackage{color}
\usepackage{soul}  % for \st{}

\newcommand{\ybco}{YBa$_2$Cu$_3$O$_7$ }

\usepackage{color}% use colored text in latex

\newcommand{\lyxmathsym}[1]{\ifmmode\begingroup\def\b@ld{bold}
  \text{\ifx\math@version\b@ld\bfseries\fi#1}\endgroup\else#1\fi}

\usepackage{amsfonts}
\usepackage{mathrsfs}

\hypersetup{colorlinks=true,urlcolor=blue,linkcolor=blue}

\begin{document}

%\title{Structural optimization of the superconducting temperature in optimally doped YBCO by many-body first principle approaches}

\title{Disentangling the role of bond lengths and orbital symmetries in controlling $T_{c}$ in optimally doped \ybco}

%\title{Strain control of superconductivity in optimally doped YBCO}
\author{Francois Jamet$^{1}$}
\email{francois.jamet@kcl.ac.uk}

\author{Cedric Weber$^{1}$}
\email{cedric.weber@kcl.ac.uk}

\author{Swagata Acharya$^{1,2}$}
\author{Dimitar Pashov$^{1}$}
\author{Mark van Schilfgaarde$^{1,3}$}
\affiliation{ $^1$King's College London, Theory and Simulation of Condensed Matter,
              The Strand, WC2R 2LS London, UK}
\affiliation{$^2$Institute for Molecules and Materials, Radboud University, NL-6525 AJ Nijmegen, The Netherlands}
\affiliation{$^3$National Renewable Energy Laboratory, Golden, Colorado 80401, USA}

\begin{abstract}

Optimally doped YBCO (YBa$_{2}$Cu$_{3}$O$_{7}$) has a high critical temperature, at 92 K. It is largely believed that Cooper pairs form in YBCO and other cuprates because of spin fluctuations, the issue and the detailed mechanism is far
from settled.  In the present work, we employ a state-of-the-art \emph{ab initio} ability to compute both the low and
high energy spin fluctuations in optimally doped YBCO. We benchmark our results against recent inelastic neutron
scattering and resonant inelastic X-ray scattering measurements. Further, we use strain as an external parameter to
modulate the spin fluctuations and superconductivity.  We disentangle the roles of Barium-apical Oxygen hybridization,
the interlayer coupling and orbital symmetries by applying an idealized strain, and also a strain with a fully relaxed
structure.  We show that shortening the distance between Cu layers is conducive for enhanced Fermi surface nesting, that
increases spin fluctuations and drives up $T_{c}$.  However, when the structure is fully relaxed electrons flow to the
d$_{z^2}$ orbital as a consequence of a shortened Ba-O bond which is detrimental for superconductivity.
\end{abstract}

\maketitle

Even while more than three decades have passed since their discovery, the origin of superconductivity in cuprates remain
highly debated.  It is largely believed that at least for several variants of cuprates it is primarily spin
fluctuations that drive superconductivity\cite{Dahm_2009,RevModPhys.84.1383}.  However, the issue is far from settled.
It is difficult to resolve because the phase diagram is very dense: small excursions in parameter space drives the
material from one phase to a new phase.  Multiple low energy scales are present and sometimes intertwined.
Theoretically, the challenge has been to build a material specific ability that incorporates all such interactions in
the right proportions which can also predict ways to disentangle their roles.  Optimally doped YBCO
(YBa$_{2}$Cu$_{3}$O$_{7}$) is one of the higher $T_{c}$ superconductors in the cuprate family.  In this work we present
a high-fidelity \emph{ab initio} theory that is designed to realize this objective, and we use it to explore how spin
fluctuations and superconductivity can be modified through strain.  In a recent ultrafast experiment \cite{Hu_2014} on
YBa$_{2}$Cu$_{3}$O$_{7}$ large amplitude distortions on apical Oxygens were induced to modulate the inter layer
coupling.  Here we apply a uniaxial strain along the \emph{c}-axis. We perform two distinct excursions under strain;
$(a)$ an ideal strain where all atoms displace in proportion to their height along the \emph{c} axis, and $(b)$ allowing
the internal coordinates to relax under the strain.  We show that strain modifies superconductivity in both cases, but
in different ways, thus highlighting how a detailed understanding of the mechanism is essential in order to predict and
control unconventional superconductivity.

High resolution inelastic neutron scattering (INS) data for spin fluctuations exists \cite{Reznik_2008,Woo_2006}, but
the data is available only up to 60\,meV.  State-of-the-art recent resonant inelastic X-ray scattering (RIXS) picks up
the signatures of bosonic fluctuations of different kinds \cite{rixs}, including those whose mechanisms are intertwined.
Thus to decipher what constitutes the primary component of the observed RIXS spectra is a challenge for both theorists
and experimentalists alike.  RIXS data has been taken from the Cu-L$_{3}$ edge for YBa$_{2}$Cu$_{3}$O$_{7}$ along the
(0,$\pi$) line \cite{rixs} with excitations observed up to 300 meV.  However, measurements could not be performed
for the important momentum region around ($\pi$, $\pi$), which likely drives superconductivity.

In the present letter, we calculate the magnetic susceptibilities and superconducting instability for  optimally doped YBCO using a new high fidelity, \emph{ab initio} approach~\cite{nickel,questaal_paper}.  For the
one-particle Green's function it combines the quasiparticle self consistent \emph{GW} (QS\emph{GW})
approximation~\cite{kotani} with CTQMC solver~\cite{hauleqmc} based dynamical mean field theory (DMFT).  This
framework~\cite{prx,Baldini} is extended by computing the local vertex from the two-particle Green's
function by DMFT~\cite{hyowon_thesis,yin}, which is combined with nonlocal bubble diagrams to construct a Bethe-Salpeter equation~\cite{swag19,prl20}.  The latter is
solved to yield the essential two-particle spin and charge susceptibilities $\chi^{d}$ and $\chi^{m}$ --- physical
observables which provide an important benchmark.  Moreover they supply ingredients needed for the Eliashberg equation,
which yields eigenvalues and eigenfunctions that describe instabilities to superconductivity.  We will denote
QS\emph{GW}\textsuperscript{\footnotesize{++}} as a shorthand for the four-tier QS\emph{GW}+DMFT+BSE+Eliashberg theory.
The numerical implementation is discussed in Pashov et al.~\cite{questaal_paper} and codes are available on the open
source electron structure suite Questaal~\cite{questaal_web}.  Some details are also given in the supplemental
material.

QS\emph{GW}\textsuperscript{\footnotesize{++}} has high fidelity because QS\emph{GW} captures non-local dynamic
correlation particularly well in the charge channel~\cite{tomc, questaal_paper}, but cannot adequately capture effects
of spin fluctuations.  DMFT does an excellent job at the latter, which are strong but mostly controlled by a local
effective interaction given by $U$ and $J$. In this letter, we have used $U=8eV$ and $J=0.7eV$ which is similar to what
is generally used for cuprates \cite{Choi2016}.  That it can well describe superconductivity has now been established in
several materials~\cite{swag19,prl20}.

In YBCO, we explore the full potential of our ability by performing rigorous bench-marking of our computed magnetic
susceptibilities (resolved in momentum and energy) against the low and high energy spectral data from INS and RIXS
respectively. We show that we reproduce all intricate structures in momentum and energy spaces observed in INS and RIXS
from our theory in a parameter free manner.  That it is possible to reproduce the RIXS spectra from the spin
susceptibility alone indicates that RIXS is measuring an excitation primarily magnetic in nature in this compound.

The superconducting glue the present theory can characterize originates from some combination of spin and charge
susceptibility~\cite{swag19,hyowon_thesis}.  That we are able to reliably recover experimental neutron and RIXS data provides some confirmation that
we have an adequate foundation to describe superconductivity of this kind.  We can use the same method to probe how spin
fluctuations and superconductivity are affected when the system is perturbed, in particular how $T_{c}$ evolves with
tensile strain.  In a prior work\cite{swag19}, this machinery was used to explain in detail how $T_{c}$ evolves with
tensile strain in Sr$_{2}$RuO$_{4}$ where it could be benchmarked against experiments.

\begin{figure}[ht!]
  \centering
  \includegraphics[scale=1]{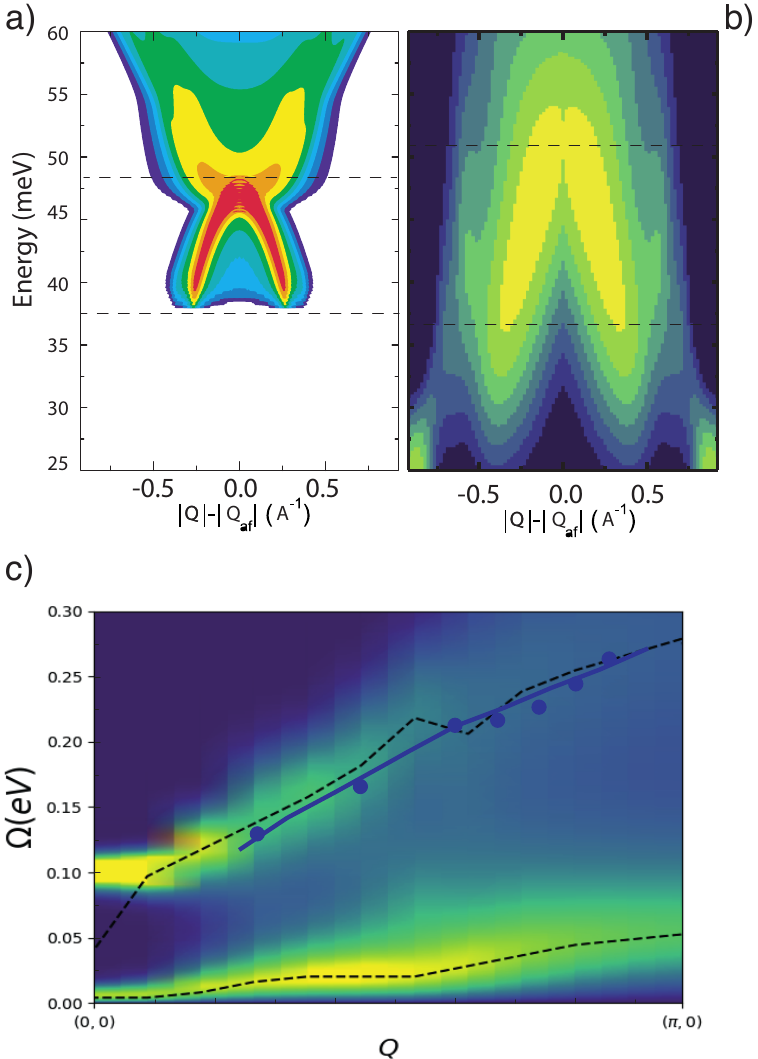}
  \caption{$(a)$ Structure factor obtained by Inelastic Neutron Scattering along the nodal line, reproduced from
    \cite{Reznik_2008}. $(b)$ Structure factor for the same energy window obtained by
    QS\emph{GW}\textsuperscript{\footnotesize{++}}.
    $(c)$:
  Heat map representing  dynamical structure factor  along the path $\Gamma$-$X$ in the Brillouin Zone computed with QS\emph{GW}\textsuperscript{\footnotesize{++}}. Dashed line indicates the position of the maxima.
  Cirle dot :RIXS experimental data shown for comparison, reproduced from \cite{Le_Tacon_2013}.}
    \label{fig:ins-rixs}
\end{figure}

Fig.~\ref{fig:ins-rixs} benchmarks the dynamical structure factor $S(q,\omega)=
({1-e^{-\beta\omega}})^{-1}\,\mathrm{Im}\,\chi^{\text{mag}}(q,\omega)$ computed by
QS\emph{GW}\textsuperscript{\footnotesize{++}} (Fig. \ref{fig:ins-rixs}$b$) against direct INS measurements of $S$
(Fig. \ref{fig:ins-rixs}$a$) in the vicinity of the antiferromagnetic point $Q_{AF}=(\pi,\pi$); and also against RIXS
which measures any bosonic excitation, including the magnetic structure factor (Fig. \ref{fig:ins-rixs}$c$).  For the
former, we can compare only up to maximum value reported, 60\,meV.  At least in this range of energy and
momentum QS\emph{GW}\textsuperscript{\footnotesize{++}} is in good qualitative and quantitative agreement.  Since the RIXS
measurement~\cite{Le_Tacon_2011} did not sample the region close to $Q_{AF}$, we compare to RIXS along the line $Q$
connecting $\Gamma$ and $(\pi,0)$.  A second branch appears at high energy.  For both low-energy and high-energy
excitations the maxima of the peaks along $Q$ line shown by the dashed line are in a very good agreement with the
experimental data (blue dots). It confirms the magnetic nature of the excitation measured in this RIXS response,
a subject of some controversy \cite{Benjamin_2014,Minola_2015,Kan_sz_Nagy_2016}.

Next we use the Eliashberg theory derived from spin and charge susceptibilities to estimate $T_c$, and investigate how
it is affected by strain. We use QS\emph{GW}\textsuperscript{\footnotesize{++}} to simulate an uniaxial strain on the
$c-$direction and study its effect on the superconducting order. The uniaxial strain is carried out by reducing the
$c-$axis up to $8\%$ with a concomitant expansion the plane. The volume change $\frac{\Delta V}{V}$ and the reduction of $c$ are related by $\frac{\Delta V}{V}{=}(1-2\nu)\frac{\Delta c}{c}$ where the Poisson ratio used is  $\nu=0.25$, which is close to
experiment \cite{Chen_2019}. For each strain, we consider two scenarios : an ideal strain where all internal
displacements are fixed to their projection along the \emph{c} axis, and another case where atoms are relaxed to the
zero-force condition.  Forces are computed within density functional theory.  We will denote these
scenarios as `SS' (for single shot) and `SO' (for structure optimized).  SO corresponds to the actual mechanical
response of YBCO subject to an $\epsilon_{33}$ tensile strain.  For convenience of presentation we define a strain
with the opposite sign of the usual definition: $\epsilon_z=-\epsilon_{33}=\frac{c_0{-}c}{c_0}$.

For both scenarios, we compute the variation of the critical temperature $T_c$ by comparing the superconducting
instability computed by solving the linearized Eliashberg equation (see Appendix).  Comparing these two scenarios
distinguishes two competing effects: on the one hand, the ideal strain changes the topology of the Fermi surface in a way that
favors superconducting order. On the other hand, subsequently allowing the internal coordinates to fully relax empties
the $d_{z^2}$ orbital, which is unfavorable for the superconducting order.
\begin{table}
\centering
\begin{tabular}{|c|cc|cc|cc|}
 \hline $\epsilon_z[\%]$ &  \multicolumn{2}{c|}{Cu$^1$-Cu$^2$[\AA]}&  \multicolumn{2}{c|}{Cu-AO[\AA]}&  \multicolumn{2}{c|}{Ba-AO[\AA]}  \\
               &  SO & SS   &  SO & SS   &  SO & SS \\ \hline
  \input{tab1.txt}
\end{tabular}
\caption{\label{tab:geo} Interlayer Cu-Cu spacing (first column), copper to apical oxygen distance (second column),
  and vertical component of AO to Ba distance (third column). We report distances for the pristine material (first row), and under
  uniaxial strain ($\epsilon_z$).   Parameters for ideal (SS) and fully relaxed (SO) structures are shown (see text).}

\end{table}

\begin{figure}[ht!]
  \centering
  \includegraphics[scale=0.4]{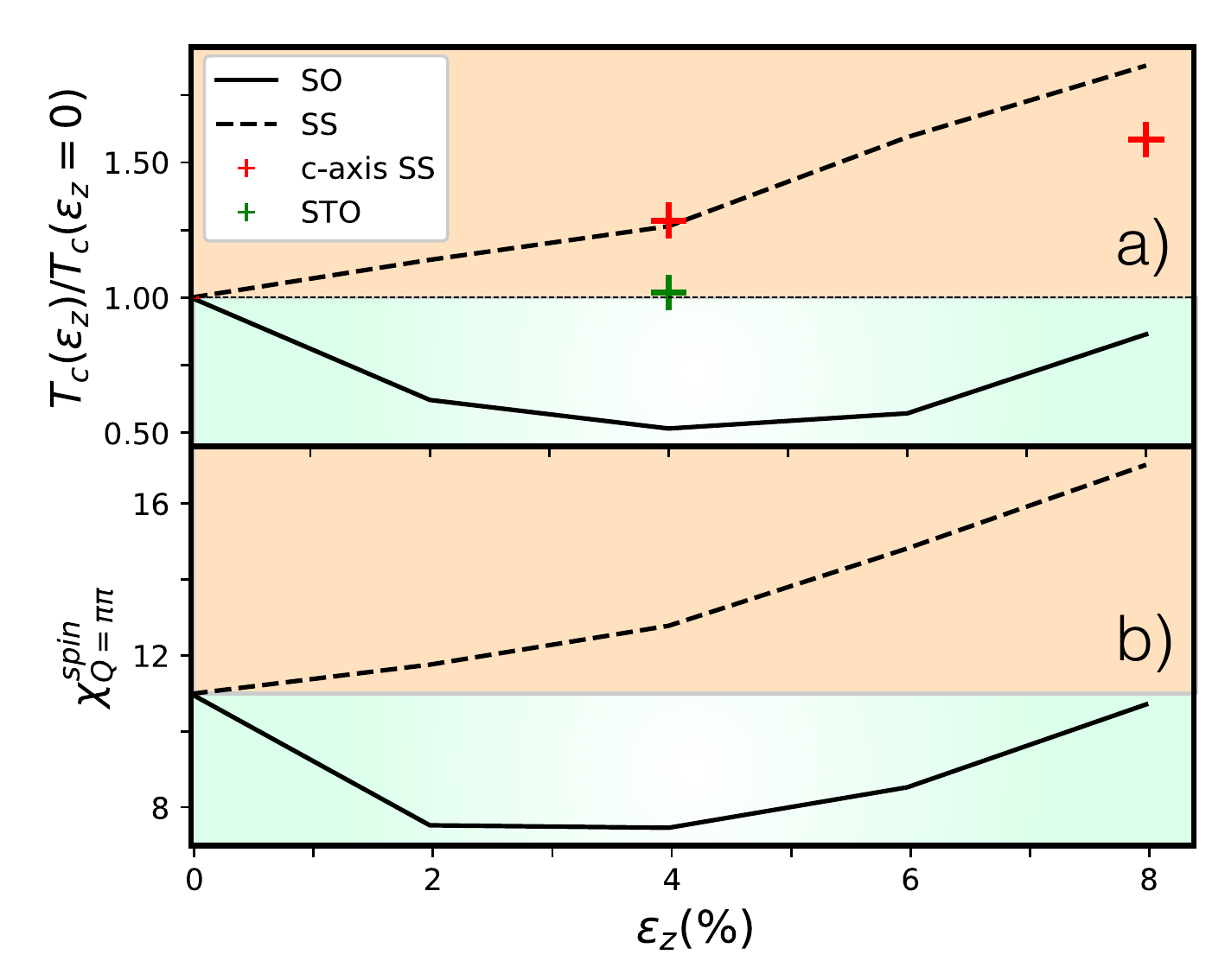}
  \caption{(Top panel) $T_{c}$ relative to the unstrained condition for the SS case (dashed line) and SO case (solid
    line).  $T_{c}$ increases in the former case, but decreases in the latter (see text).  The red crosses show the SS
    case assuming a Poisson ratio of 0 (basal plane is kept frozen).
    This shows that the dominant effect is the change in coupling along the \emph{c} axis.  The green cross corresponds to a compression of 2\% in the basal plane in SS case.
    Panel $(b)$ shows the corresponding evolution of the  static  magnetic susceptibility at
    $Q=(\pi,\pi)$. It close correspondence with $T_c$ indicates that spin fluctuations are the dominant contributor to the
    superconducting instability. Panel $(c)$: evolution of the Stoner factor $s=1/(1-\alpha)$ where $\alpha$ is the
    highest eigenvalue of $\chi_0 \Gamma$ in $\chi = \chi_0 / (1-\chi_0 \Gamma)$ }.
    \label{fig:observables}
\end{figure}

Fig.~\ref{fig:observables} shows these results in more detail.  $T_c$ is greatly increased in the SS scenario, and the
middle and bottom panels show how $T_c$ is correlative to magnetic susceptibility at $(\pi,\pi)$.  Both the spin
fluctuation and the superconductivity instability show a similar trend which confirms that the spin fluctuation are an
essential contributor of the superconductivity.  We can also separate the contributions from stretching the \emph{c}
axis from the contributions by reductions in the basal plane by varying the Poisson ratio.  In one case we used $\nu$=0
which freezes the lattice vector in the basal plane (red symbols in Fig.~\ref{fig:observables}).  $T_c$ changes only
marginally with $\nu$, which indicate that the main effect are coming from the reduction of the $c$-axis.  In another
scenario, we expand $a$ and $b$ axes about $2\%$ to match \ybco epitaxially on an STO substrate (green
symbol).

On the  $(0,0){-}(\pi,\pi)$ line, the Cu $d_{x^2-y^2}$ in the two planes of {YBa$_2$Cu$_3$O$_{7-\delta}$}
  couple weakly through the apical O, splitting these otherwise degenerate states into a bond-antibond pair.
  The Fermi surface connected with these orbitals splits into two sheets.

\begin{figure}[ht!]
  \centering \includegraphics[scale=0.23]{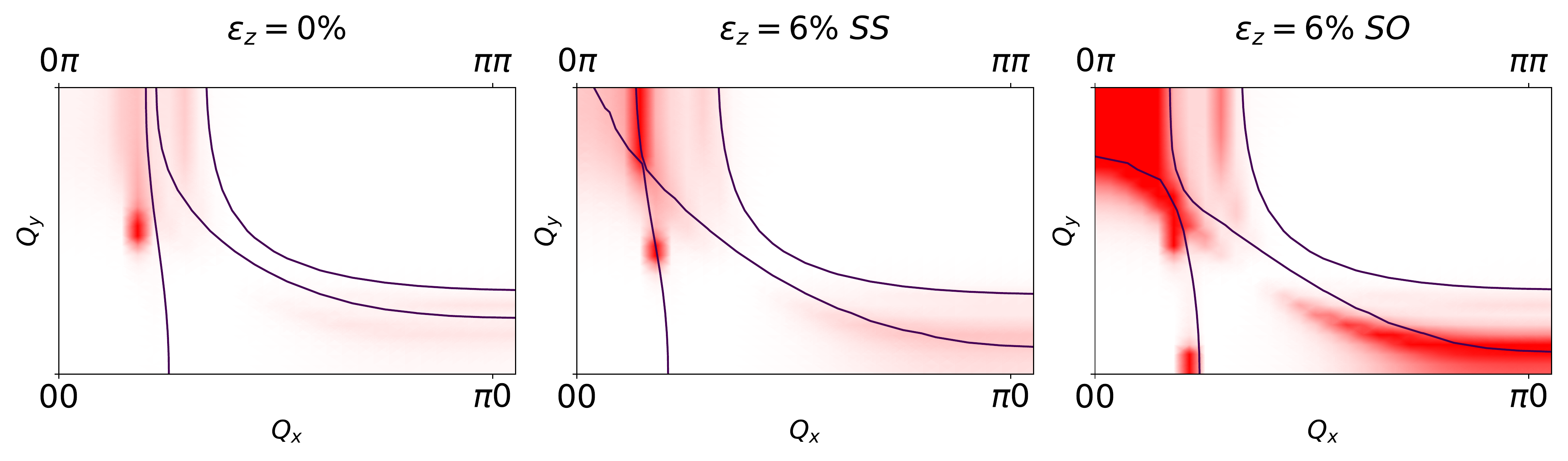} \caption{Evolution of the Fermi surface against applied strain, as
    computed from  QS\emph{GW}\textsuperscript{\footnotesize{++}}. The
    Fermi surface (full lines) is composed of a vertical line originating from the CuO chain, and other lines from the
    bilayer coupling as explained in the text.   Strain increases the separation between antibonding and bonding Fermi sheets
    for both $(a)$ ideal strain (SS) and $(b)$ fully relaxed strain (SO).
    SS and SO differ in the degree that $d_{x^2-y^2}$ and $d_{z^2}$ are coupled, as discussed in the text.
    This is manifest by the colorbar, which shows the off-diagonal component $|G(Q,\omega{=0})_{{z^2},{x^2-y^2}}|$,
    an indicator of the hybridization between $d_{z^2}$ and $d_{x^2-y^2}$ on the Fermi surface.}
  \label{fig:fs}
\end{figure}

\begin{figure}[ht!]
  \centering
  \includegraphics[scale=0.5]{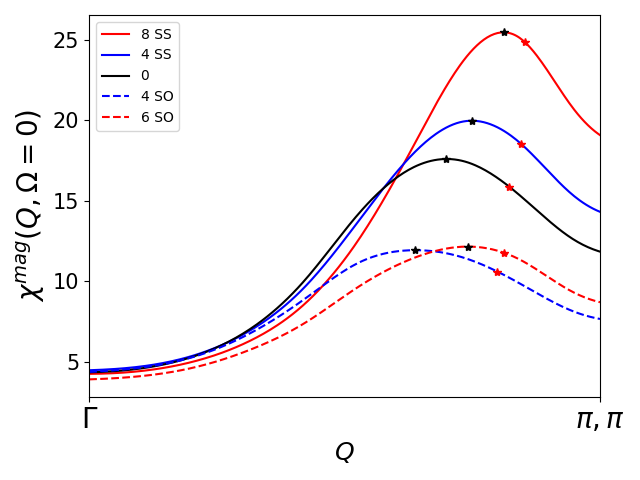}
  \caption{ Static magnetic
      susceptibility $\chi(q,\omega{=}0)$ on the $(0,0){-}(\pi,\pi)$ line, for pristine YBCO (black line), and YBCO
    subject to SS and SO kinds of strain.  SS and SO induce opposite effects on both the peak position and amplitude
    $\chi(q,\omega{=}0)$.  In the SS case $T_{c}$ increase both because $\chi(q,\omega{=}0)$ increases and the peak
    shifts closer to $Q_{AF}$.}
  \label{fig:magsus}
\end{figure}

As shown in Fig.~\ref{fig:fs}, the Fermi surface is formed of three bands. The two curved lines correspond
  to the bonding and antibonding $d_{x^2-y^2}$ bands noted above. The interlayer hybridization is strongest at the two
  antinodal points where either $Q_x{=}\pi$ or $Q_{y}{=}\pi$.  As strain is applied, the interlayer distance decreases
  (Table~\ref{tab:geo}).  This increases the interlayer hybridization, which further splits the bonding and antibonding
  $d_{x^2-y^2}$ Fermi surfaces.  The antibonding surface becomes flatter and thus more square.  Making the arc more
  square improves the nesting of  momentum transfer $Q{=}(\pi,\pi)$ for electrons living at the antinodal point. In d-wave superconductivity, it increases the attractive interaction \cite{Bickers_1987,Weber_2012}, i.e   works constructively for $d$-wave superconductivity.  As shown in Fig.~\ref{fig:magsus}, not
  only the magnetic susceptibility increases but the nesting vector moves closer to $Q_{AF}$ where the magnetic
  susceptibility is maximum. These two effects cumulatively explain the large enhancement of $T_c$ observed in
  the SS scenario.

In SO scenario, we observe a similar splitting of  the bonding and antibonding $d_{x^2-y^2}$ Fermi surfaces, but with an important difference.  In SO case, the $d_{x^2-y^2}$  hybridize  with $d_{z^2}$ .  This hybridization increases because relaxation reduces the Ba-AO vertical distance, e.g. by $25\%$ when the \emph{c} axis is reduced by $4\%$ (Table~\ref{tab:geo}).  As a consequence, the AO environment is changed which affects the Cu $d_{z^2}$ orbital.  In the unstrained case, $d_{z^2}$ sits at $-$1.48\,eV below the Fermi level, and it is marginally changed in the SS case (for $\epsilon_z{=}4\%$ it resides at $-$1.41\,eV) while in the SO case, $d_{z^2}$ is pushed closer to the Fermi level ($-$1.18\,eV).  The Fermi surface mainly composed of $d_{x^2-y^2}$ state becomes strongly hybridized with $d_{z^2}$.  This is apparent from colorbar in Fig.~\ref{fig:fs}.  The two orbitals hybridize close the antinodal point which is known to be unfavorable to \emph{d}-wave superconductivity \cite{PhysRevB.85.064501,Matt_2018,PhysRevLett.105.057003}. Indeed, not only almost filled band, as $d_{z^2}$ is unfavorable for $T_c$ but opposite spin nearest-neighbor coupling to form Cooper pair is less favorable when two orbitals are active at the Fermi level. When $d_{z^2}$ orbital are not included in correlated subspace and in the Eliashberg equation, this detructive effect disappears.

To recap, in the idealized scenario compression of the planes increases the interlayer hybridization which enhances
Fermi surface nesting and, hence, $T_c$ However, with a proper relaxation (SO), the Ba-AO bond length decreases
dramatically to force out-of-plane contributions to the planar physics and changes the orbital components of the Fermi
surface. This contribution works destructively for $T_c$.

To conclude, we have first shown that QS\emph{GW}\textsuperscript{\footnotesize{++}} is able to predict magnetic
fluctuations in \ybco with high fidelity.  Using this technique to explore the parameter space when \ybco is subject to
strain, we find that $T_c$ can be dramatically altered, but how it is altered depends on the details of the
displacements.  Subject to an ideal strain with no internal relaxation, $T_c$ is increased owing to enhanced interlayer
hybridization, which changes the shape of the Fermi surface and makes nesting more favorable.  However, in a more
realistic scenario, the Fermi surface also suffers from a competing $d_{z^2}$ hybridization which  is detrimental to $T_c$.

\section*{Acknowledgments}
CW acknowledges insightful and stimulating discussions with Antoine Georges.
This work was supported by the Simons Many-Electron collaboration.  CW was supported by grant EP/R02992X/1 from the UK Engineering and Physical Sciences Research Council (EPSRC). F. J. are supported by the EPSRC Centre for Doctoral Training in Cross-Disciplinary Approaches to Non-Equilibrium Systems (CANES, EP/L015854/1). F.J is supported by the Simons Many-Electron Collaboration.
For computational resources, we thank PRACE for awarding us access to SuperMUC at GCS@LRZ, Germany and Irene-Rome hosted by TGCC, France and Cambridge Tier-2 system operated by the University of Cambridge Research Computing Service (www.hpc.cam.ac.uk) funded by EPSRC Tier-2 capital Grant No. EP/P020259/1 and ARCHER UK National Supercomputing Service.

\section*{Supplementary material}

The calculation in the unstrained case was perfomed using crystal struture reported in \cite{PhysRevB.41.1863}. Paramagnetic DMFT is combined with non magnetic QS\emph{GW} via local projection on Cu 3\emph{d} on the Cu  augmentation  spheres  to  form  the  correlated  subspace. DMFT loop was performed using CTQMC impurity solver \cite{hauleqmc} and \cite{Seth2016274}. The two particle susceptibility needed in BSE for the magnetic susceptibility was computed using Exact Diagonalisation impurity solver with 6 bath sites on a mesh of 50 bosonic frequencies and 500 fermionic frequencies. A benchmark was done with CTQMC solver to check the accuracy of the hybridization fit. The Eliashberg equation was solved using an impurity susceptibility computed with CTQMC impurity solver\cite{Acharya_2019}.

\bibliography{dmft}

\end{document}

%% file: tab1.txt
0.0 & 3.39 & 3.39 & 2.30 & 2.30 & 0.30 & 0.30 \\ 
\hline2.0 & 3.38 & 3.33 & 2.21 & 2.26 & 0.26 & 0.29 \\ 
\hline4.0 & 3.32 & 3.28 & 2.15 & 2.22 & 0.22 & 0.29 \\ 
\hline